\documentclass{aa}  
\pdfoutput=1
\usepackage{tabularx}
\usepackage{array}
\usepackage{graphicx}
\usepackage{txfonts}
\usepackage[colorlinks,citecolor=blue]{hyperref}
\usepackage{orcidlink}

\begin{document}

   \title{Scaling up global kinetic models of pulsar magnetospheres using a hybrid force-free-PIC numerical approach}

   \titlerunning{Scaling up kinetic models of pulsar magnetospheres}

   \author{Adrien Soudais
          \inst{1}\orcidlink{0000-0002-8599-8847}
          Benoît Cerutti\inst{1}\orcidlink{0000-0001-6295-596X}
          \and Ioannis Contopoulos\inst{2}\orcidlink{0000-0001-6890-4143}}

   \institute{Univ. Grenoble Alpes, CNRS, IPAG, 38000 Grenoble, France\\
              \email{adrien.soudais@univ-grenoble-alpes.fr}
         \and
            Research Center for Astronomy and Applied Mathematics, Academy of Athens, GR 11527 Athens, Greece
             }

\date{Received 04 April 2024 / Accepted 12 June 2024}

\abstract
{The particle-in-cell approach has proven effective at modeling neutron star and black hole magnetospheres from first principles, but global simulations are plagued with an unrealistically small separation between the scales where microphysics operates and the system-size scales due to limited numerical resources. A legitimate concern is whether the scale separation currently achieved is large enough, such that results can be safely extrapolated to realistic scales.}
{In this work, our aim is to explore the effect of scaling physical parameters up, and to check whether salient features uncovered by pure kinetic models at smaller scales are still valid, with a special emphasis on particle acceleration and high-energy radiation emitted beyond the light cylinder.}
{To reach this objective, we develop a new hybrid numerical scheme coupling the ideal force-free and the particle-in-cell methods, to optimize the numerical cost of global models. We propose a domain decomposition of the magnetosphere based on the magnetic field topology using the flux function. The force-free model is enforced along open field lines while the particle-in-cell model is restricted to the reconnecting field line region.}
{As a proof of concept, this new hybrid model is applied to simulate a weak millisecond pulsar magnetosphere with realistic scales using high-resolution axisymmetric simulations. Magnetospheric features reported by previous kinetic models are recovered, and strong synchrotron radiation above 100MeV consistent with the Fermi-LAT gamma-ray pulsar population is successfully reproduced.}
{This work further consolidates the shining reconnecting current sheet scenario as the origin of the gamma-ray emission in pulsars, as well as firmly establishes pulsar magnetospheres as at least TeV particle accelerators.}
   

\maketitle

\section{Introduction}

Magnetospheres forming around neutron stars and black holes, or simply referred to as relativistic magnetospheres in the following, are involved in some of the most extreme high-energy astrophysical phenomena, such as pulsars \citep{2017SSRv..207..111C, 2022ARA&A..60..495P}, magnetars flares and outbursts \citep{1995MNRAS.275..255T, 2017ARA&A..55..261K}, relativistic jets from supermassive black holes \citep{1977MNRAS.179..433B, 2019ARA&A..57..467B}, and possibly fast radio bursts \citep{2010vaoa.conf..129P, 2020ApJ...897....1L} and precursor emission to compact object mergers \citep{2019A&A...622A.161C, 2020ApJ...893L...6M}. Elaborating a quantitative and predictive model of relativistic magnetospheres turned out to be an outstanding endeavor at the forefront of modern physics, and our current understanding of such systems is at best incomplete. The fact that these regions concentrate large energy densities into various forms (i.e., gravitational, rotational, electromagnetic), and the interplay between plasma physics, general relativity and relativistic (quantum) electrodynamics dramatically increases the degree of complexity of the problem that becomes analytically untractable.

In the quest for a comprehensive theory of relativistic magnetospheres, the aligned rotator has historically played the role of a Rosetta stone in this field. Some of the most important developments were obtained in the framework of force-free electrodynamics, a degenerate form of relativistic magnetohydrodynamic (MHD) in which the evolution of the plasma is solely governed by the Lorentz force \citep{2002luml.conf..381B, 2002MNRAS.336..759K}. Thus, all of the other forces including inertia or gas pressure are irrelevant in this regime. This is a valid assumption almost everywhere within the magnetosphere where the field strength is high and the plasma density is low. In this limit, the structure of the magnetosphere is then a solution of the relativistic Grad-Shafranov equation \citep{1973ApJ...182..951S, 1973ApJ...180L.133M}, and an exact solution can be derived for a magnetic monopole \citep{1973ApJ...180L.133M, 1977MNRAS.179..433B}. Extending these results to a more complex magnetic field topology, even to an aligned magnetic dipole, has never been achieved by analytical means.

Instead, an approximate numerical model must be used. The force-free solution of the aligned dipole recovers essential magnetospheric features \citep{1999ApJ...511..351C, 2006MNRAS.367...19K, 2006ApJ...648L..51S, 2006MNRAS.368.1055T, 2006MNRAS.368L..30M, 2012MNRAS.423.1416P, 2016MNRAS.455.4267C}, in particular the overall hybrid magnetic topology composed of closed and open field lines and the formation of an equatorial current layer separating both magnetic hemispheres beyond the light-cylinder radius, where co-rotation with the star is superluminal \citep{1969ApJ...157..869G, 1969Natur.223..277M}. Thereby, the force-free framework provides a realistic description for the morphology of relativistic magnetospheres in the limit where plasma is abundant and ultramagnetized everywhere. It also captures well how the star spins down under the effect of magnetic torques. However, the main shortcoming of this approach is that it cannot capture dissipation (see, however \citealt{2008arXiv0802.1716G} and \citealt{2012ApJ...746...60L} for a dissipative force-free formulation), and more importantly non-thermal particle acceleration and radiation. It is therefore impossible to pin down the emitting regions and to constrain the model with observations without using ad-hoc prescriptions, leading to large theoretical uncertainties \citep{2010ApJ...715.1282B, 2014ApJ...793...97K, 2019ApJ...874..166C}.

Observations of rotation-powered pulsars in the gamma-ray range demonstrate that the magnetosphere accelerates particles with a disconcerting efficiency \citep{2010ApJS..187..460A, 2013ApJS..208...17A, 2023ApJ...958..191S}. The recent detections of TeV pulsed emission in the Crab and Vela pulsars indicate that electrons are accelerated at least up to 20TeV \citep{2016A&A...585A.133A, 2023NatAs...7.1341H}, which represents a sizeable fraction of the full polar-cap potential drop. Acceleration of hadrons in the magnetosphere is also an open issue of great interest for understanding the origin of cosmic rays (e.g., \citealt{1969PhRvL..22..728G, 2000ApJ...533L.123B, 2003ApJ...589..871A, 2012ApJ...750..118F}). To address these outstanding questions and to connect the physical model to observations, an ab-initio plasma model is required, taking into account basic plasma processes at microscopic (i.e., kinetic) scales, pair production, the radiation-reaction force, and general relativistic effects. In this respect, the particle-in-cell (PIC) approach has proven effective at integrating all of the above physical ingredients necessary to model relativistic magnetospheres from first principles \citep{2014ApJ...785L..33P, 2014ApJ...795L..22C, 2015MNRAS.448..606C, 2015MNRAS.449.2759B, 2015ApJ...801L..19P, 2020A&A...635A.138G, 2020ApJ...889...69C, 2022ApJ...939...42H, 2023ApJ...958L...9B, 2023arXiv230904834C, 2023ApJ...943..105H, 2024arXiv240102908T}. Global PIC simulations of pulsar magnetospheres show robust features such as a sizeable dissipation rate of the outflowing Poynting flux via magnetic reconnection of open field lines beyond the light cylinder. This energy is efficiently channelled into non-thermal particle acceleration and energetic radiation, possibly explaining the observed high-energy radiation \citep{2016MNRAS.457.2401C, 2018ApJ...855...94P, 2018ApJ...857...44K}. At the base of the open field lines near the star, the spark-gap dynamics and the putative radio emission are also well reproduced by local PIC models \citep{2013MNRAS.429...20T, 2020PhRvL.124x5101P, 2021ApJ...919L...4C}.

In spite of these impressive developments, these results must be interpreted with caution because global PIC simulations are plagued with an unrealistically small separation between the scales where microphysics operates, and the system-size (i.e., light-cylinder) scales. A legitimate concern is whether the scale separation achieved by current global PIC simulations is large enough such that results can be safely extrapolated to realistic scales. Thus, capturing the polar-cap physics and the light-cylinder reconnection physics with full scales in the same simulation seems impossible, even in a far future. In this work, we make the choice to sacrifice the polar-cap microphysics to focus our numerical resources on the light-cylinder scales, where most of the dissipation and particle acceleration take place \citep{2020A&A...642A.204C, 2023ApJ...943..105H}. Our main objective is to scale physical parameters up, and to check whether salient features uncovered by pure kinetic models at smaller scales still hold strong. To this end, we develop a new hybrid numerical scheme to take advantage of the complementarity between the force-free and the PIC approaches. We propose a domain decomposition based on the magnetic field topology of the system, where open field lines are described by an ideal force-free model while the reconnecting region is described by a PIC model. As a proof of concept, we apply this method to the aligned pulsar magnetosphere with the aim at reaching realistic scales of a weak millisecond pulsar, just above the limit of the observed \emph{Fermi}-LAT pulsar population. In the next section (Sect.~\ref{sect::scales}), we present the relevant physical scales needed to model a pulsar magnetosphere. In Section~\ref{sect::hybrid}, we describe the new hybrid force-free-PIC numerical scheme. This approach is first tested against the monopole solution (Sect.~\ref{sect::monopole}). It is then applied to the aligned dipole with high-resolution simulations, where we explore the effect of scaling up physical parameters, from previously achieved scales in full PIC simulations, to more realistic scales (Sect.~\ref{sect::dipole}). We summarize this work and we discuss future applications and developments of this new hybrid approach in Section~\ref{sect::conclusion}.

\section{Scale separation}\label{sect::scales}

Capturing all relevant physical scales in pulsar magnetospheres is an outstanding challenge for simulations. In the following discussion, we will consider, as our baseline, a millisecond pulsar with a rotation period $P=1$ms and a surface dipolar magnetic field $B_{\star}=10^7$G, corresponding to the weakest observable gamma-ray pulsars of spindown power $L\approx 4.8\times 10^{33}$erg/s as reported by the \emph{Fermi}-LAT \citep{2010ApJS..187..460A, 2013ApJS..208...17A, 2023ApJ...958..191S}. It also represents a realistic gamma-ray pulsar with the smallest scale separation that we strive to capture in this work as a proof of principles (see Sect.~\ref{sect::dipole}).

The system-size macroscopic scales are set by the neutron star radius, $r_{\star}\sim 10^6$cm, and the light-cylinder radius,
\begin{equation}
R_{\rm LC}=\frac{cP}{2\pi}\approx 5r_{\star} \left(\frac{P}{1\rm{ms}}\right).
\end{equation}
The light-cylinder radius sets the scale at which the plasma streams away from the star along open fields in the form of a relativistic wind. Magnetic reconnection starts operating at the base of the wind current layer, known as the ``Y-point'' (e.g., \citealt{2003ApJ...598..446U, 2024MNRAS.527L.127C}), and proceeds at all radii beyond this point. Energetic synchrotron emission is produced within a few light-cylinder radii because of the steep decay of the magnetic field strength with radius (\citealt{2016MNRAS.457.2401C}; Cerutti et al. in preparation), and thus sets the largest relevant scale of the problem in this work.

While the macroscopic scales are well constrained, defining the microscopic scales is more involved because plasma physics and QED effects must be taken into consideration. The electronic plasma skindepth at the star surface sets the minimal relevant scale of the problem. Assuming that the plasma density is a multiple of the critical co-rotation density, $n_{\star}=\kappa n^{\star}_{\rm GJ}=\kappa B_{\star}/2\pi e R_{\rm LC}$ \citep{1965JGR....70.4951H, 1969ApJ...157..869G}, where $\kappa$ is the plasma multiplicity, and $e$ is the electron charge, yields
\begin{eqnarray}
d^{\star}_{\rm e}&=&\sqrt{\frac{\gamma m_{\rm e}c^2}{4\pi n_{\star} e^2}}\nonumber \\
&\approx& 2\left(\frac{\gamma}{10^0}\right)^{1/2}\left(\frac{\kappa}{10^2}\right)^{-1/2}\left(\frac{P}{1\rm{ms}}\right)^{1/2}\left(\frac{B_{\star}}{10^7\rm{G}}\right)^{-1/2} \rm{cm},
\label{eq::skindepth}
\end{eqnarray}
where $\gamma$ is the particle Lorentz factor and $m_{\rm e}$ is the electron mass, so that $d^{\star}_{\rm e}/r_{\star}=2\times 10^{-6}$ for our reference pulsar case. The situation may still look hopeless at this stage. However, particles at the neutron star surface are not at rest. Instead, they are believed to move at ultra-relativistic speeds ($\gamma\gg 1$), hereby reducing the above scale separation. A primary beam of particles is extracted and accelerated by the surface electric field induced by the rotation of the star \citep{1971ApJ...164..529S, 1975ApJ...196...51R}. At best, the particles will undergo the full vacuum potential drop across the polar cap of the star. The size of the polar cap is defined between the magnetic axis and the first field line fully contained within the light cylinder. For an aligned dipole in vacuum, its angular size is $\sin\theta_{\rm pc}=\sqrt{r_{\star}/R_{\rm LC}}$. Using the corotation surface electric field, $\mathbf{E}=-(\boldsymbol{\Omega}\times\mathbf{r})\times\mathbf{B}/c$, we can derive the potential drop across the polar cap as
\begin{equation}
\Phi_{\rm pc}=\frac{\mu\Omega^2}{c^2},
\label{eq::phi_pc}
\end{equation}
where $\Omega=2\pi/P$ is the angular velocity of the star, and $\mu=B_{\star}r^3_{\star}$ its magnetic moment. An electron experiencing the full potential drop will acquire a Lorentz factor given by
\begin{equation}
\gamma_{\rm pc}=\frac{e\Phi_{\rm pc}}{m_{\rm e}c^2}\approx 2.6\times 10^8 \left(\frac{B_{\star}}{10^7\rm{G}}\right)\left(\frac{P}{1\rm{ms}}\right)^{-2}.
\label{eq::gamma_pc}
\end{equation}
This calculation provides an upper limit, it is not a realistic estimate of the particle Lorentz factor in the bulk of the flow, because the presence of the plasma screens in great part the accelerating electric field (parallel to the magnetic field), and because of pair production. The main channel for pair production near the star surface is believed to be by magnetic conversion, meaning that a gamma-ray photon is annihilated by virtual photons from the magnetic field to produce a pair. Quantum electrodynamics predicts that pair production occurs if \citep{1966RvMP...38..626E, 2006RPPh...69.2631H}
\begin{equation}
\chi\equiv\epsilon b\gtrsim 0.1,
\label{eq::chi}
\end{equation}
where $\epsilon=\hbar\nu/m_{\rm e}c^2$ is the dimensionless photon energy, and $b=\tilde{B}_{\perp}/B_{\rm QED}$ is the effective perpendicular magnetic field (see Eq.~\ref{eq::bperp} below) normalized to the critical quantum field $B_{\rm QED}=m^2_{\rm e}c^3/\hbar e\approx 4.4\times 10^{13}$G. If electrons radiate via synchrotron-curvature radiation as usually assumed, the critical photon energy is given by \citep{1970RvMP...42..237B}
\begin{equation}
\epsilon_{\rm c}=\frac{3}{2}b\gamma^2.
\label{eq::epsilon}
\end{equation}
Combining Eq.~(\ref{eq::chi}) with Eq.~(\ref{eq::epsilon}) provides an estimate for the threshold electron Lorentz factor capable of producing new pairs,
\begin{equation}
\gamma_{\rm th}=\sqrt{\frac{1}{15 b^2}}\approx 10^6\left(\frac{B_{\star}}{10^7\rm{G}}\right)^{-1},
\label{eq::gamma_th}
\end{equation}
while the created pair shares the absorbed photon momentum, such that the Lorentz factor of the secondary pairs may be estimated as
\begin{equation}
\gamma_{\rm s}=\frac{\epsilon}{2}=\frac{1}{20b}\approx 2.2\times 10^5 \left(\frac{B_{\star}}{10^7\rm{G}}\right)^{-1}.
\end{equation}
Assuming that pair creation is effective at producing a high-multiplicity plasma in the magnetosphere ($\kappa\gg 1$), it sets the relevant scale of the particle Lorentz factor to be considered in estimating the plasma skindepth scale in Eq.~(\ref{eq::skindepth}), yielding $d^{\rm s}_{\rm e}=d^{\star}_{\rm e}(\gamma_{\rm s})\approx 10^3\rm{cm}=10^{-3}r_{\star}$ for our reference pulsar. One can realize from these order of magnitude calculations that a weaker magnetic field ($B_{\star}\lesssim 10^7$G) would prevent pair formation as the threshold energy scale would be larger than the full potential drop.

At the light cylinder, a thin current sheet forms whose thickness, $\delta$, is governed by the local plasma skindepth scale, meaning that $\delta \sim d^{\rm LC}_{\rm e}$. At this stage, efficient particle acceleration proceeds via magnetic reconnection, increasing on average the particle Lorentz factor to a scale governed by the plasma magnetization parameter,
\begin{equation}
\gamma_{\rm LC}\sim \sigma_{\rm LC}=\frac{B^2_{\rm LC}}{4\pi \Gamma_{\rm LC}\kappa n^{\rm LC}_{\rm GJ}m_{\rm e}c^2}= \frac{\gamma_{\rm pc}}{2\Gamma_{\rm LC}\kappa},
\label{eq::sigmalc}
\end{equation}
where $B_{\rm LC}\sim B_{\star}(r_{\star}/R_{\rm LC})^3=\mu\Omega^3/c^3$, $n^{\rm LC}_{\rm GJ}=\Omega B_{\rm LC}/2\pi e c$ is the Goldreich-Julian plasma density at the light cylinder, and $\Gamma_{\rm LC}$ is the pulsar wind bulk Lorentz factor at its base. Assuming that these energetic particles set the local skindepth scale yields
\begin{equation}
\delta \sim d^{\rm LC}_{\rm e}(\gamma_{\rm LC})=\sqrt{\frac{\gamma_{\rm LC}m_{\rm e}c^2}{4\pi \kappa \Gamma_{\rm LC} n^{\rm LC}_{\rm GJ}e^2}}=\frac{R_{\rm LC}}{2\Gamma_{\rm LC}\kappa},
\end{equation}
\begin{equation}
\delta\sim 2.4\times 10^{4}\left(\frac{\Gamma_{\rm LC}}{10^0}\right)^{-1}\left(\frac{\kappa}{10^2}\right)^{-1}\left(\frac{P}{1\rm{ms}}\right)\rm{cm},
\end{equation}
thus, $\delta/r_{\star}\sim 2\times 10^{-2}$ or $\delta/R_{\rm LC}\sim 5\times 10^{-3}$ for these fiducial parameters. A similar estimate can be obtained from Amp\`ere law \citep{1990ApJ...349..538C, 2001ApJ...547..437L, 2017A&A...607A.134C} (see also, \citealt{2013A&A...550A.101A, 2014ApJ...780....3U}, for other estimates).

Radiative cooling sets another energy scale in the problem at which the accelerating electric force is balanced by the radiation-reaction force, $eE_{\parallel}\sim f_{\rm rad}$. The latter effectively acts as a continuous drag force (in the classical regime, i.e., $\gamma b \ll 1$), opposite to the particle direction of motion, whose magnitude in the ultrarelativistic regime ($\gamma\gg 1$) can be estimated as $f_{\rm rad}\approx(2/3)r^2_{\rm e}\gamma^2 \tilde{B}_{\perp}^2$, where $r_{\rm e}$ is the classical radius of the electron. The fiducial radiation-reaction-limited electron Lorentz factor is then given by
\begin{equation}
\gamma_{\rm rad}=\sqrt{\frac{3eE_{\parallel}}{2 r^2_{\rm e} \tilde{B}^2_{\perp}}}\approx 3\times 10^4 \left(\frac{E_{\parallel}}{\tilde{B}_{\perp}}\right)^{1/2}\left(\frac{\tilde{B}_{\perp}}{10^7\rm{G}}\right)^{-1/2},
\end{equation}
or $\gamma^{\rm LC}_{\rm rad}\approx 3.3\times 10^5$ at the light cylinder for the fiducial pulsar parameters, assuming $E_{\parallel}=\tilde{B}_{\perp}$ and $\tilde{B}_{\perp}=B_{\rm LC}$.

In summary, modeling the weakest gamma-ray pulsar must satisfy at the very least the following scale separation, in terms of spatial quantities normalized to $r_{\star}$,
\begin{equation}
\frac{d^{\rm s}_{\rm e}}{r_{\star}}\sim 10^{-3} \ll \frac{\delta}{r_{\star}} \sim 2.5\times 10^{-2} \ll 1 < \frac{R_{\rm LC}}{r_{\star}}=5,
\end{equation}
and in terms of energy scales normalized to $\gamma_{\rm pc}$
\begin{equation}
\frac{\gamma_{\rm s}}{\gamma_{\rm pc}}\sim 8.5\times10^{-4}< \frac{\gamma^{\rm LC}_{\rm rad}}{\gamma_{\rm pc}}\sim 10^{-3} <\frac{\gamma_{\rm th}}{\gamma_{\rm pc}}\sim 4\times10^{-3}\ll 1.
\label{eq::scale_separation}
\end{equation}
This exercise reveals that in such system $\gamma_{\rm s}$, $\gamma_{\rm rad}$ and $\gamma_{\rm th}$ are very close together meaning that the energy range of the particle spectrum may be quite narrow. This property is in agreement with a new important feature highlighted by the most recent analysis of the \emph{Fermi}-LAT pulsars, which shows that the observed spectral energy distributions is narrower with decreasing spindown, nearly reaching a spectrum consistent with a monoenergetic particle population for the weakest pulsars of spindown $L\sim 10^{33}$erg/s \citep{2023ApJ...958..191S}. Hereby, even though the estimates presented in this section represent by no means a rigorous calculation, it provides an approximate assessment of the minimum computational needs to approach a realistic systems using a kinetic approach. Our conclusion is that such weak pulsar could in principle be captured with full scales with current computational facilities, provided that extreme resolutions are used in a 2D axisymmetric system, as for instance in the recent works by \citet{2022ApJ...939...42H, 2023ApJ...958L...9B}.

\section{Hybrid force-free-PIC approach}\label{sect::hybrid}

In order to fulfil the minimum scale separation requirements highlighted in the previous section, a large number of grid cells must be used, and hence a large computational power is needed. To alleviate some of this numerical cost, we develop a new hybrid approach specially designed for aligned magnetospheres, coupling time-dependent force-free electrodynamic and the PIC approaches in the same simulation. This idea relies on the fact that gamma-ray pulsar magnetospheres are mostly very close to a dissipationless force-free state because of efficient pair creation (i.e., $\kappa\gg1$). Using the expensive full PIC technique with a large number of particles per cell everywhere in the magnetosphere may appear as excessive, while a numerically cheaper force-free description would be sufficient. In contrast, dissipative regions (e.g., electrostatic gaps, current sheets) should remain on microscopic scales (again, if $\kappa\gg 1$), and so are the scales at which the PIC approach is really needed. Therefore, it appears desirable to combine these two techniques, but a new scheme must be implemented to ensure that both approaches can coexist harmoniously together.

\begin{figure}
    \centering
    \includegraphics{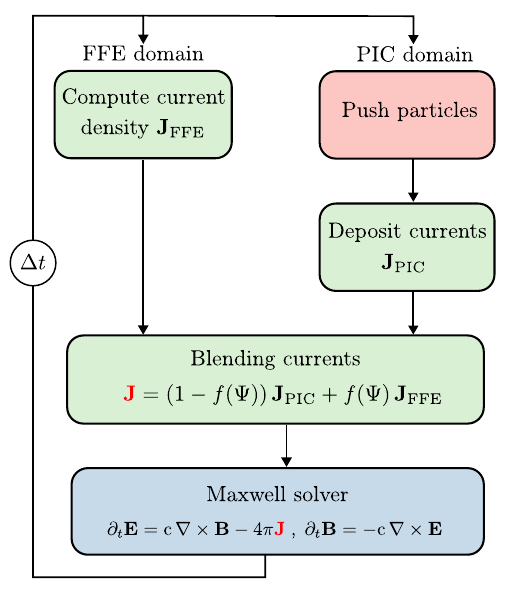}
    \caption{This diagram presents the new hybrid numerical scheme proposed in this work during a time integration cycle. This method is meant to blend the force-free electrodynamic (FFE) and the full kinetic (PIC) approaches within the same numerical framework using a domain decomposition. The coupling is achieved via the electric current densities, $\mathbf{J}$, using a blending function $f$ which solely depends on the magnetic flux function $\Psi$.}
    \label{fig::hybrid_step}
\end{figure}

In this work, we develop a new force-free module implemented in the {\tt Zeltron} code, an explicit relativistic PIC code \citep{2013ApJ...770..147C} used in the past for global full PIC simulations of pulsar magnetospheres (e.g., \citealt{2015MNRAS.448..606C}). In the PIC approach, the plasma is modeled by point-like superparticles --or simply referred as particles in the following-- representing a very large number of physical particles with identical mass-to-charge ratio following the same path in phase space \citep{1991ppcs.book.....B}. Their evolution is governed by the Lorentz force, and the radiation-reaction force in this pulsar-specific context, as
\begin{equation}
\frac{\mathrm{d}\mathbf{u}}{\mathrm{d}t} = \frac{q}{mc}\left(\mathbf{E}+\frac{\mathbf{u}\times\mathbf{B}}{\gamma}\right)+\mathbf{g},
\label{eq::motion}
\end{equation}
where $\mathbf{u}=\gamma\mathbf{v}/c$ is the dimensionless particle 4-momentum vector, and $q$ is the particle electric charge, and where
\begin{equation}
\mathbf{g}\approx\frac{2}{3}r^2_{\rm e}\left[\left(\mathbf{E}+\boldsymbol{\beta}\times\mathbf{B}\right)\times\mathbf{B}+\left(\boldsymbol{\beta}\cdot\mathbf{E}\right)\mathbf{E}\right]-\frac{2}{3}r^2_{\rm e}\gamma^2\tilde{B}^2_{\perp}\boldsymbol{\beta},
\end{equation}
is the radiation reaction force following the \citet{1971ctf..book.....L} formulation, and $\boldsymbol{\beta}=\mathbf{v}/c$, and where
\begin{equation}
\tilde{B}_{\perp}=\sqrt{\left(\mathbf{E}+\boldsymbol{\beta}\times\mathbf{B}\right)^2-\left(\boldsymbol{\beta}\cdot\mathbf{E}\right)^2}\label{eq::bperp},
\end{equation}
is the effective perpendicular magnetic field \citep{2016MNRAS.457.2401C}. Particle trajectories are evolved in time with a modified Boris pusher to incorporate the radiation reaction force to the Lorentz force. The current density carried by each particle is deposited onto the numerical grid where the electromagnetic fields are defined. Summing over all particles and over all species, the total current $\mathbf{J}$ reconstructed on the grid is then used to evolve the time-dependent Maxwell equations,
\begin{equation}
\frac{\partial\mathbf{B}}{\partial t} = -c\,\boldsymbol{\nabla}\times\mathbf{E} \label{eq::faraday}
\end{equation}
\begin{equation}
\frac{\partial\mathbf{E}}{\partial t} = c\,\boldsymbol{\nabla}\times\mathbf{B}-4\pi\mathbf{J},
\label{eq::ampere}
\end{equation}
allowing to close the PIC loop performed at each time step (Fig.~\ref{fig::hybrid_step}). The field integration step follows a standard finite-difference time-domain method that involves a staggered mesh \citep{1966ITAP...14..302Y}. 

In general, each simulation cell is filled with a large number of particles, and thus pushing the particles and depositing the currents largely dominate the overall computing time. These two steps typically take one order of magnitude longer than evolving the fields alone. This is where the benefits of force-free electrodynamics come into play. The force-free condition is given by 
\begin{equation}
F_{\mu\nu}I^{\nu}=0,
\end{equation}
where $F_{\mu\nu}$ is the electromagnetic tensor and $I^{\nu}$ is the 4-current, which translates into the following conditions
\begin{equation}
\rho\mathbf{E}+\frac{\mathbf{J}\times\mathbf{B}}{c}=\mathbf{0},
\label{eq::ff}
\end{equation}
using the spatial components, where $\rho=\boldsymbol{\nabla}\cdot\mathbf{E}/4\pi$ is the charge density, and
\begin{equation}
\mathbf{E}\cdot\mathbf{J}=0,
\label{eq::EdotJ}
\end{equation}
using the temporal component. Eq.~(\ref{eq::ff}) also leads to the condition 
\begin{equation}
\mathbf{E}\cdot\mathbf{B}=0.
\label{eq::EdotB}
\end{equation}
Combining Eqs.~(\ref{eq::ff})-(\ref{eq::EdotB}) with Maxwell equations Eqs.~(\ref{eq::faraday})-(\ref{eq::ampere}), one can express the total current density solely from the fields in the form of an effective Ohm's law, as \citep{1999astro.ph..2288G, 2002luml.conf..381B}
\begin{equation}
    \mathbf{J} = \frac{c}{4\pi}\boldsymbol{\nabla}\cdot\mathbf{E}\left(\frac{\mathbf{E}\times\mathbf{B}}{B^2}\right)+\frac{c}{4\pi}\Bigl(\mathbf{B}\cdot\left(\boldsymbol{\nabla}\times\mathbf{B}\right)-\mathbf{E}\cdot\left(\boldsymbol{\nabla}\times\mathbf{E}\right)\Bigr)\frac{\mathbf{B}}{B^2},
    \label{eq::JFFE}
\end{equation}
which can be understood as the sum of a current flowing across field lines, $\mathbf{J_{\perp}}$, and a component flowing along field lines, $\mathbf{J_{\parallel}}$. Instead of pushing a large number of particles and depositing the current on the grid, the current is derived in a single step using Eq.~(\ref{eq::JFFE}), which saves on computing time and memory. Other than this, the Maxwell solver is unchanged whether the current is computed via the particles or via the fields. Thus, the coupling between both approaches is done via the current.

To implement the force-free module, we follow a similar numerical procedure as in \citet{2006ApJ...648L..51S}, which consists in solely computing the perpendicular force-free current, $\mathbf{J}_{\perp}$. The parallel component, $J_{\parallel}$, prevents the formation of an electric field component along the magnetic field at all times, $\mathbf{E}\cdot\mathbf{B}=0$, but computing $J_{\parallel}$ is a numerically delicate and cumbersome endeavor on a staggered mesh (see, however, \citealt{2012MNRAS.423.1416P}). Therefore, the electric field is renormalized at each time step to enforce the force-free condition,
\begin{equation}
\mathbf{E}\leftarrow\mathbf{E}-\left(\mathbf{E}\cdot\mathbf{B}\right)\frac{\mathbf{B}}{B^2}.
\end{equation}
The code also checks at every time step that the condition $E^2<B^2$ is always satisfied, so that if it is not the case, the electric field is once again renormalized as
\begin{equation}
\mathbf{E}\leftarrow\sqrt{\frac{B^2}{E^2}}\mathbf{E}.
\end{equation}
Computing the force-free current and renormalizing the electric field directly on the staggered mesh is delicate, because field components defined at different locations on the cell and times are mixed, leading to a disruption of the second-order accuracy of the scheme. Instead, we interpolate all quantities into the same grid point to compute the current and the electric field, and then interpolate these back onto the staggered grid. The interpolation scheme is based on an arithmetic average between neighboring points.

To blend the two approaches, the computing box is decomposed into pure force-free and pure PIC domains. In axisymmetric systems, using the magnetic flux function is a natural choice to make this division. This scalar field is constant along poloidal field lines which are themselves equipotential surfaces, it is therefore a physically motivated criterion which is also numerically convenient to tag individual field lines. In ordinary spherical coordinates ($r,\theta$), the magnetic flux function is given by
\begin{equation}
\Psi=\frac{1}{2\pi}\iint \mathbf{B}\cdot d\mathbf{S}=\int_0^\theta B_{\mathbf r}r^2\sin\theta^{\prime}d\theta^{\prime}.
\label{eq::psi}
\end{equation}
Once the boundary is set, a thin transition layer is implemented between neighbouring field lines where the currents are blended to ensure a better continuity between the PIC and the force-free currents and to minimize numerical artefacts, such as spurious reflections or accumulation of charges. A simple linear interpolation scheme is sufficient for this purpose, so that the full hybrid current is expressed as
\begin{equation}
\mathbf{J}=\Bigl(1-f\left(\Psi\right)\Bigr)\,\mathbf{J}_{\rm PIC}+f\left(\Psi\right)\,\mathbf{J}_{\rm FFE},
\label{eq::blending}
\end{equation}
where $\mathbf{J}_{\rm PIC}$ is the current density reconstructed from the particles, $\mathbf{J}_{\rm FFE}$ is the current from the force-free condition (Eq.~\ref{eq::JFFE}), and $f$ is the blending function. This hybrid current is then injected into Maxwell-Amp\`ere law (Eq.~\ref{eq::ampere}) to evolve the field in the whole simulation domain. The formulation of Eq.~(\ref{eq::blending}) implies that both schemes are computing the current in the transition layer (i.e., particles are present everywhere within this layer, but they are of course absent in the pure force-free domain). Considering a single transition layer in the domain, two distinct fluxes must be chosen, $\Psi_0<\Psi_1$, so that the blending function is
\begin{eqnarray}
f\left(\Psi\right)&=&1,~\text{if}~\Psi<\Psi_0,\rightarrow \text{pure force-free, no particles},\\
f\left(\Psi\right)&=&\frac{\Psi_1-\Psi}{\Psi_1-\Psi_0},~\text{if}~\Psi_0<\Psi<\Psi_1,\rightarrow \text{transition layer},\label{eq::transition}\\
f\left(\Psi\right)&=&0,~\text{if}~\Psi>\Psi_1,\rightarrow \text{pure PIC}.
\end{eqnarray}
The magnetic flux function is computed at each time step so that the blending function is updated accordingly. In the context of pulsar magnetospheres where the magnetic field lines are frozen into the neutron star surface, this decomposition is fixed on the neutron star surface and then follows the evolution of the magnetic field topology in time in the rest of the domain. In this sense, the domain decomposition proposed in this work is not static but dynamically evolves along with the magnetospheric activity. Figure~\ref{fig::hybrid_step} graphically summarizes the proposed numerical scheme over a time iteration.

\section{Monopole}\label{sect::monopole}

The hybrid scheme is first put to the test on the modeling of a steady-state force-free monopolar magnetosphere (i.e., a single magnetic monopole). Although unphysical, studying this configuration is of great interest: it admits an exact analytical solution and it is free of dissipation --there is no gap or current sheet (as opposed to a split-monopole)-- so that spurious numerical effects in the scheme can be quantified, and it represents a good model for the asymptotic magnetospheric structure of pulsar open field lines. All things considered, it sets the best conditions to validate our new approach.

In this setup, the magnetic flux function is given by
\begin{equation}
\Psi_{\rm M}=\Psi_{\star}\left(1-\cos\theta\right),
\end{equation}
where $\Psi_{\star}=B_{\star}r^2_{\star}$. An exact solution to the Grad-Shafranov equation yields the toroidal magnetic field component,
\citep{1973ApJ...180L.133M}
\begin{equation}
B_{\phi}=-\frac{\Psi_{\star}}{R_{\rm LC}}\frac{\sin\theta}{r},
\end{equation}
if $\boldsymbol{\Omega}\cdot\mathbf{B}>0$ in the northern hemisphere (and $\boldsymbol{\Omega}\cdot\mathbf{B}<0$ in the southern hemisphere). In this solution, the electric field is purely along $\theta$ and matches the toroidal magnetic field, $E_{\theta}=B_{\phi}$. The electric current is purely radial and is given by
\begin{equation}
\mathbf{J_{\rm M}}=\frac{c}{4\pi}\boldsymbol{\nabla}\times\mathbf{B}=-J^{\star}_{\rm GJ}\left(\frac{r_{\star}}{r}\right)^2\cos\theta~\mathbf{e_{\rm r}},
\label{eq::michel_current}
\end{equation}
where $J^{\star}_{\rm GJ}=\Omega B_{\star}/2\pi$ corresponds to the fiducial Goldreich-Julian current, and the spindown power associated to the electromagnetic torque is
\begin{equation}
L_{\rm M}=\frac{c}{4\pi}\iint\left(\mathbf{E}\times\mathbf{B}\right)\cdot d\mathbf{S}=\frac{c}{2}\int_{-1}^{1} r^2 B^2_{\phi}d\cos\theta=\frac{2\Psi^2_{\star}\Omega^2}{3c}.
\label{eq::michel_spindown}
\end{equation}

The goal of this first numerical experiment is to recover these properties using our hybrid scheme. To this end, we perform a set of hybrid simulations in 2D axisymmetric spherical coordinates and explore the effect of resolution on the results. The domain size explored here is composed of $1024^2$, $2048^2$ and $4096^2$ cells. The grid spacing is logarithmic along the radial direction and constant along the $\theta$ direction. The domain size ranges from the star surface, $r_{\rm min}=r_{\star}$, up to $r_{\rm max}=(10/3) R_{\rm LC}$, where $R_{\rm LC}=5r_{\star}$, and covers the full range in $\theta\in[0,\pi]$. The magnetosphere is initialized in vacuum with a pure radial magnetic field, $B=B_{\star}(r_{\star}/r)^2$. The surface magnetic field strength is fixed at $B_{\star}=5\times10^5$G, which is sufficiently intense in this experiment to achieve a quasi force-free state in the PIC domain. The role of the field strength and scale separation is thoroughly investigated in the next section. 

Corotation of the field lines with the star at the inner boundary is guaranteed by fixing the poloidal electric field to the ideal solution $\mathbf{E_{\star}}=-(\boldsymbol{\Omega}\times\mathbf{r_{\star}})\times\mathbf{B_{\star}}/c$. This procedure sets the magnetosphere into rotation. An absorbing layer starting at $r_{\rm abs}=3 R_{\rm LC}$ progressively damps all outgoing electromagnetic waves and removes the particles, if any \citep{2015MNRAS.448..606C}. In this experiment, the northern hemisphere is modeled in full PIC, while the southern hemisphere is captured with the force-free model. This choice is well suited for a comparison between both approaches thanks to the expected symmetry between both hemispheres, and hence, any asymmetry or imbalance can be tracked easily. In this setup, we did not notice any difference with or without a transition layer for the two highest resolutions, so that here, we only show the runs performed with a sharp transition at $\theta=\pi/2$, or $\Psi=\Psi_{\star}$. In the PIC domain, electron-positron pairs are continuously injected with a high multiplicity, $\kappa_{\rm inj}=10$, at the star surface to ensure a force-free state, and pair creation is not allowed elsewhere in the domain. The pairs are injected within the first layer of cells above the surface, and there are no ions at this stage.

\begin{figure}[!]
    \centering
    \includegraphics[scale=0.5]{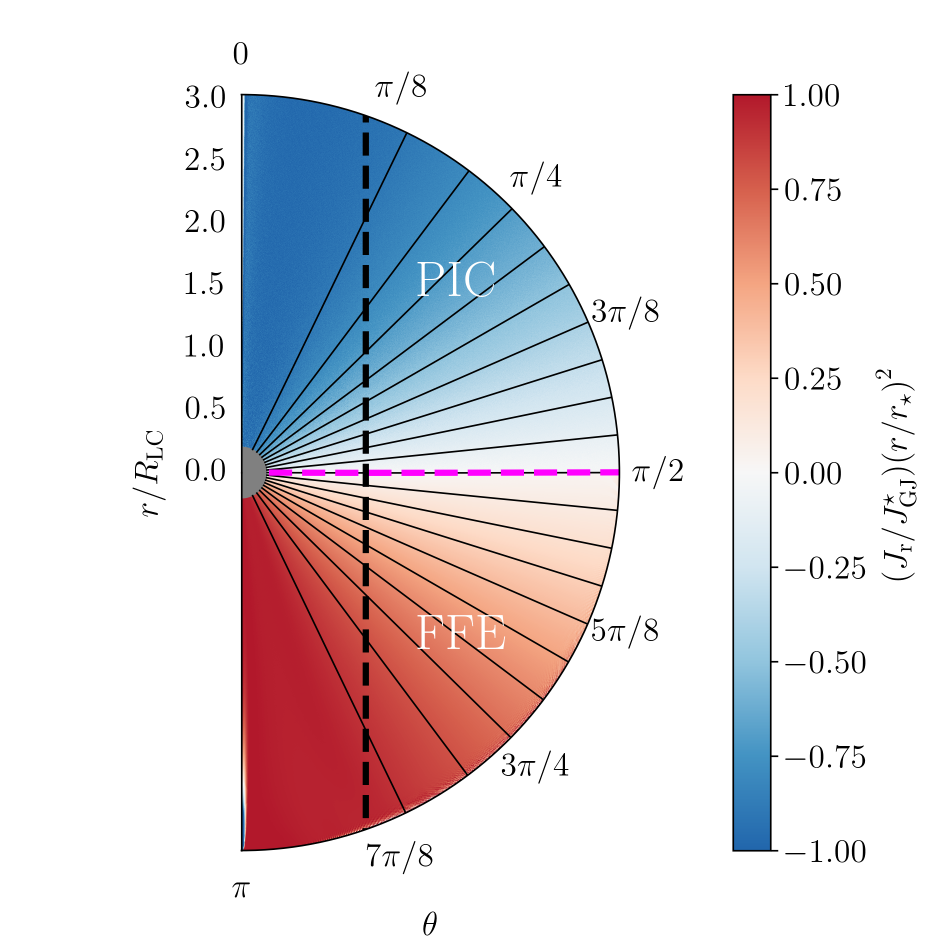}
    \includegraphics[scale=0.5]{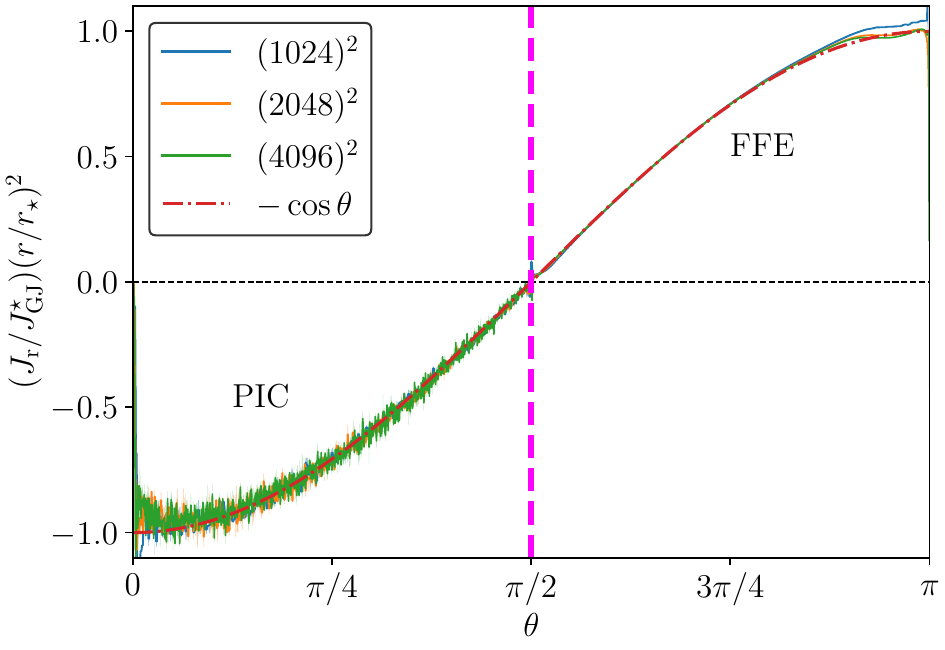}
    \includegraphics[scale=0.5]{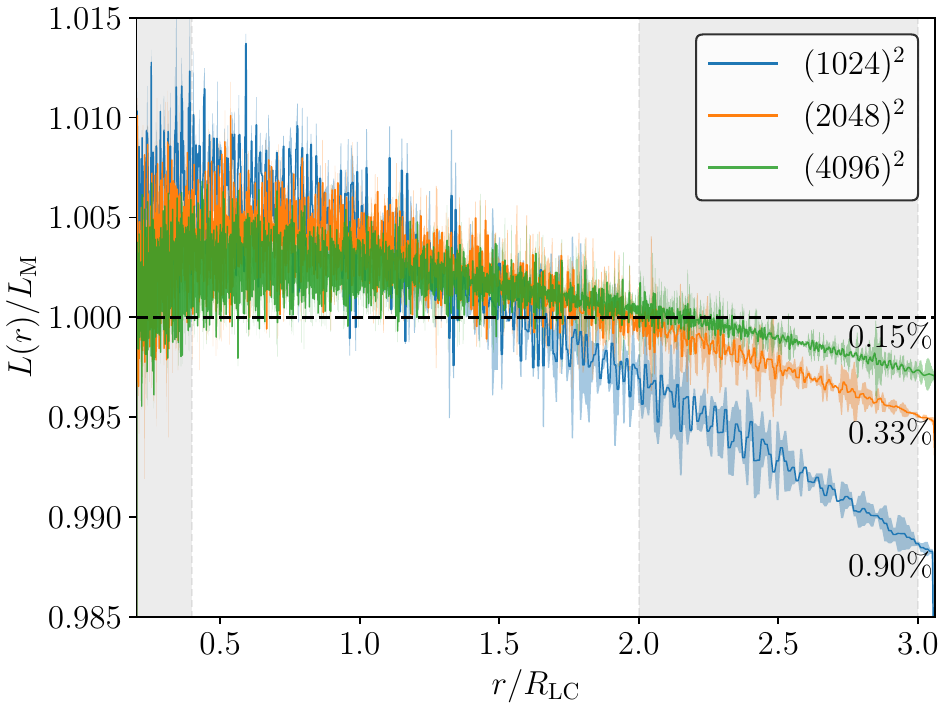}
    \caption{Simulated monopolar magnetosphere using the hybrid force-free-PIC scheme. The division between the force-free and the PIC domains is the equatorial plane (dashed magenta line in the top and middle panels). Top: Spatial distribution of the radial current density normalized to the fiducial current density, $J^{\star}_{\rm GJ}$, and compensated by $(r/r_{\star})^2$ for visualization purposes, for the $4096^2$ cells run. Solid lines show the poloidal magnetic field lines. Middle: Cross-section of the current density profile at $r=1.5 R_{\rm LC}$ as a function of the numerical resolution, and comparison with the exact profile (red dashed-dotted line, Eq.~\ref{eq::michel_current}). Bottom: Radial profiles of the Poynting flux, $L(r)$, normalized to the monopole solution, $L_{\rm M}$ (Eq.~\ref{eq::michel_spindown}), for all three resolutions. Percentages correspond to the amount of numerical dissipation for each run, computed between the flux averaged in the gray areas.}
    \label{fig::monopole}
\end{figure}

Figure~\ref{fig::monopole}, top panel, shows the state of the simulation for the highest numerical resolution, obtained after about two spin periods. A steady state is reached and the solution is numerically stable. We recover the correct variations of the toroidal field and current densities. The latter follows the predicted cosine evolution in both hemispheres without apparent mismatch at the interface between both domains at the equator, except for the lowest resolution simulation where a small glitch is visible (middle panel in Fig.~\ref{fig::monopole}). This result demonstrates that both hemispheres are well balanced, meaning that there is no net accumulation of charges at the boundary or anywhere in the magnetosphere, the outgoing current in the force-free region is perfectly matched by the incoming current in the PIC region, such that there is no net current overall. The only noticeable difference in the current profile is the high level of noise in the PIC domain as opposed to the force-free region. It is in part due to the shot noise associated with the finite number of particles, but also to fast plasma oscillations. The bottom panel in Figure~\ref{fig::monopole} zooms in on the radial evolution of the Poynting flux. The solution predicts that this quantity should be perfectly conserved at all radii (see, Eq.~\ref{eq::michel_spindown}), so that any deviation offers a way to measure the amount of numerical dissipation as a function of resolution. The rate of dissipation is computed as
\begin{equation}
\epsilon=\frac{\left|L_{\rm in}-L_{\rm out}\right|}{L_{\rm out}} \label{eq::dissipation_monopole}
\end{equation}
where $L_{\rm in}$ is the Poynting flux averaged in the $r\in[r_{\star},2r_{\star}]$ interval, and $L_{\rm out}$ is averaged over the $r\in[2R_{\rm LC},3R_{\rm LC}]$ interval (gray intervals in Fig.~\ref{fig::monopole}, bottom panel). We observe numerical convergence and a deviation smaller than $\epsilon<1\%$ for the $1024^2$ run, down to $\epsilon\sim 0.15\%$ for the highest resolution. We note that numerical dissipation mainly originates from the force-free hemisphere. This rate is about two orders of magnitude smaller than the amount of physical dissipation reported in the next section and in previous global kinetic model of pulsar magnetospheres over a similar radial scale (e.g., \citealt{2014ApJ...785L..33P, 2015MNRAS.448..606C, 2015MNRAS.449.2759B, 2020A&A...642A.204C, 2023ApJ...943..105H}). Thus, the hybrid scheme does not add significantly more artificial dissipation to the simulated system, at least under the ideal condition of a force-free monopole.

\section{Aligned dipole}\label{sect::dipole}

The hybrid model is now applied to the aligned rotating dipole using 2D axisymmetric simulations, aiming at scaling relevant physical scales up to the fiducial weak gamma-ray pulsars introduced in Sect.~\ref{sect::scales}.

\subsection{Numerical setup}

\begin{table}
\setlength{\arrayrulewidth}{0.5mm}
\[\def\arraystretch{1.35}
\begin{tabularx}{0.5\textwidth}{>{\raggedright\arraybackslash}X >{\raggedright\arraybackslash}X}
    \hline
     Physical parameters & Values  \\
     \hline
     Neutron star radius & $r_{\star}=10\;\mathrm{km}$\\
     Spin period & $P=1\,\mathrm{ms}$\\
     Surface magnetic field & $B_{\star}=10^{7}\;\mathrm{G}$\\
     Spindown power & $L_0=4.8\times10^{33}\rm{erg/s}$\\
     Polar cap energy & $\gamma_{\rm pc}=2.6\times 10^8$\\
     Threshold energy & $\gamma_{\rm th}=10^6$ \\
     Secondary pairs energy & $\gamma_{\rm s}=2.2\times10^5$\\
     Radiation-reaction energy & $\gamma^{\rm LC}_{\rm rad}=3.3\times10^5$ \\
     Electron plasma skindepth & $d_{\rm e}^{\rm s}=10^3\mathrm{cm}$\\
     Plasma timescale & $\omega^{-1}_{\rm pe}=3.3\times 10^{-8}\text{s}$\\
     Synchrotron cooling time & $t^{\rm LC}_{\rm syn}=3.6\times 10^{-7}\text{s}$\\
    \hline
     Numerical parameters & Values  \\
     \hline
     \# grid cells ($r,\theta$) & $8192\times8192$\\
     Proton to electron mass ratio & $m_{\rm p}/m_{\rm e}=1836$\\
     Time step & $\Delta t=4.3\times 10^{-9}$\text{s}\\
     Highest spatial resolution & $\Delta r=3.4\times 10^2$\text{cm}\\
     FFE domain I boundaries & $\Psi_{\rm min}=0,~\Psi_1=0.9\Psi_{\rm pc}$\\
     PIC domain boundaries & $\Psi_0=0.85\Psi_{\rm pc},~\Psi_3=2.4\Psi_{\rm pc}$\\
     FFE domain II boundaries & $\Psi_2=2.3\Psi_{\rm pc},~\Psi_{\rm max}=\mu/r_{\star}$\\
     \hline
\end{tabularx}
\]
\caption{Physical and numerical parameters for the reference weak gamma-ray pulsar simulation using the hybrid force-free-PIC scheme. Energy scales refer to electron Lorentz factors.}
\label{table::sim_params}
\end{table}

\begin{figure*}[h]
    \centering
    \includegraphics[width=18cm]{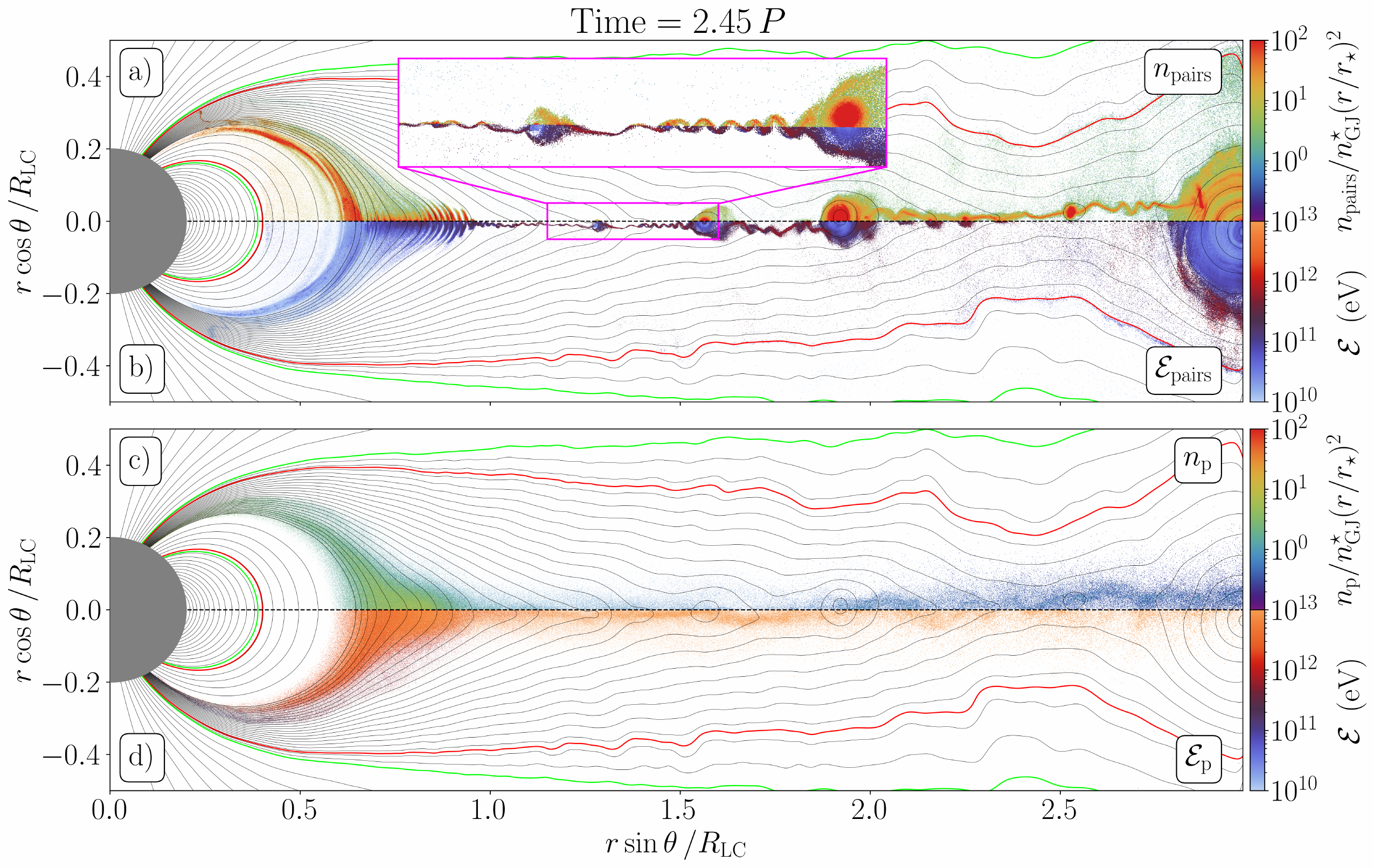}
    \caption{Global hybrid force-free-PIC model of the aligned magnetosphere of a $B_{\star}=10^7$G surface magnetic field neutron star with a $P=1$ms spin period at simulated time $t=2.45$ms. Black solid lines show the poloidal magnetic field lines (i.e., isocontour of the magnetic flux function $\Psi$), green and red solid lines show the transition from the force-free regions to the PIC region (i.e., $\Psi_0$, $\Psi_1$, $\Psi_2$, $\Psi_3$, see Table~\ref{table::sim_params}). Panel a): Map of the pairs density normalized by $n^{\star}_{\rm GJ}$ and compensated by $(r/r_{\star})^2$. Panel b): Map of the mean pair energy, $\mathcal{E}=\gamma m_{\rm e}c^2$. Panels c) and d): same as panels a) and b) for the protons. The panels do not show the full domain but instead are focused on the PIC domain where the current sheet forms. A zoomed-in view on a reconnection site is shown in panel a). Notice how thin the reconnection layer is.}
    \label{fig::densities}
\end{figure*}

The properties of the simulation domain are nearly identical as for the monopole configuration presented in the previous section (i.e., domain size, grid spacing and geometry, boundary conditions, pulsar spin period). Here, we focus on the highest resolution simulations that are composed of $8192^2$ cells. As for the monopolar magnetosphere, the box is initially empty of plasma and a dipolar field fills the full domain, and corotation is enforced at the inner boundary by setting the ideal electric field. The initial magnetic flux function is
\begin{equation}
\Psi_{\rm D}=\frac{\mu\sin^2\theta}{r}.
\label{eq::flux_dipole}
\end{equation}
There is no exact analytical solution in the force-free limit for an aligned dipole, and thus results are compared with previous force-free and full PIC simulations. In the PIC domain, plasma injection is done differently than in the monopolar case. First, the primary beam of particles is generated by the extraction from the crust of the star of electrons and protons with a realistic mass ratio, $m_{\rm p}=1836 m_{\rm e}$. This processes can be faithfully reproduced by maintaining the Goldreich-Julian plasma density at the inner boundary and at every time step. The plasma is injected in corotation with the star, but with no momentum along the poloidal field lines. The surface electric field induced by the stellar rotation then pulls and accelerates charges in the magnetosphere where pair creation can then operate. The pair production process is highly simplified, it is based on a constant energy threshold criterion of the parent electron or positron everywhere in the simulation, including in the wind zone. If the lepton Lorentz factor is larger than $\gamma_{\rm th}$, then a new pair is created with a Lorentz factor $\gamma_{\rm s}$ at the same position and along the same direction of motion as the parent particle. This procedure is similar to the heuristic models of pair production performed in previous studies \citep{2015ApJ...801L..19P, 2020ApJ...889...69C, 2021ApJ...919L...4C}. It provides an effective way to inject pairs in the magnetosphere with the relevant energy scale, but this model should not be viewed as very realistic. Pair production in millisecond pulsars remains a puzzle that we do not wish to address in this work. In the force-free domain, there is no other condition to consider apart from corotation at the neutron star surface.

Our choice for an efficient force-free-PIC domain decomposition in this setup is educated from the magnetic topology revealed by previous studies. For a plasma-filled magnetosphere, the main loci of dissipation are the equatorial current sheet and at its base around the Y-point region near the light cylinder, and to a lesser extent the separatrix current layers forming between the closed and open field line boundary that connect the equatorial current sheet to the star polar caps. These are also the sites for the putative gamma-ray emission in observed pulsars that we seek to reproduce in this work. Hereby, a natural choice is to limit the PIC domain to the equatorial and separatrix regions. The boundaries of the domain should be chosen with care. It is particularly important that the PIC domain extends from the star surface all the way to the absorbing boundary. If the chosen field line separating the PIC and the force-free domains reconnects inside the box, particles can no longer escape and the current layer is abruptly disrupted, leading to artificial dissipation. On the other hand, it would fail the purpose to choose a wide PIC domain. We found that a good compromise is to use the following magnetic flux function boundaries:
\begin{itemize}
    \item Force-free domain I, $0<\Psi<0.9\Psi_{\rm pc}$. This region corresponds to the open field lines connected to the star polar caps.\\
    \item PIC domain, $0.85\Psi_{\rm pc}<\Psi<2.4\Psi_{\rm pc}$. The domain encompasses the equatorial and separatrix current layers.\\
    \item Force-free domain II, $2.3\Psi_{\rm pc}<\Psi<\Psi_{\rm max}$. It is inside the closed field line regions deep inside the light cylinder.
\end{itemize}
In the above, $\Psi_{\rm max}=\mu/r_{\star}$ is the maximum of the flux function that is located at the star surface in the equatorial plane, and $\Psi_{\rm pc}=\mu/R_{\rm LC}$ is the magnetic flux crossing the light cylinder for a dipole in vacuum. This latter quantity is a good proxy for the amount of flux of open magnetic field lines even for the force-free dipole. There is a $\Delta\Psi=0.1\Psi_{\rm pc}$-thick transition layer between the PIC and force-free domains. However, during the transient phase, we found that the transition layer between the force-free domain I and the PIC domain can sometimes partially prevent field line opening, so that it is best to set a sharp transition in this phase. This transition layer is enforced once the initial transient has passed.

The star radius is fixed at $r_{\star}=10^6$cm, and the light-cylinder radius to $R_{\rm LC}=5\times10^6$cm corresponding to a $P=1$ms spin period. The surface magnetic field of the reference production run is fixed at $B_{\star}=10^7$G, giving $B_{\rm LC}=\mu/R^3_{\rm LC}\approx 8\times 10^4$G at the light cylinder, and a force-free spindown power $L_0=\mu^2\Omega^4/c^3=4.8\times 10^{33}$erg/s. The magnetic field strength then sets the relevant energy scales given in Sect.~\ref{sect::scales} as, $\gamma_{\rm pc}=2.6\times 10^8$, $\gamma_{\rm th}=10^6$, $\gamma_{\rm s}=2.2\times 10^5$ and $\gamma_{\rm rad}=3\times 10^4$. The Larmor radius and frequency scales are not resolved near the surface of the star, but they are irrelevant since the particles stream along magnetic field lines with negligible perpendicular momentum due to the strong radiative cooling. However, the plasma skindepth and frequency scales are resolved by $d^{\rm s}_{\rm e}/\Delta r\approx 3$ and $\omega^{-1}_{\rm pe}/\Delta t\approx 8$ at the surface where the conditions are numerically the most demanding, and they are well resolved elsewhere in the magnetosphere. The Larmor scales are also resolved in the outer parts of the magnetosphere, in particular in the equatorial current layer where these scales become relevant. The Larmor radius of secondary pairs at the light cylinder, $\rho_{\rm s}$, divided by the local grid spacing, $R_{\rm LC}\Delta r/r_{\star}$, is $\rho_{\rm s}/(R_{\rm LC}\Delta r/r_{\star})=(\gamma_{\rm s}m_{\rm e}c^2/eB_{\rm LC})/(R_{\rm LC}\Delta r/r_{\star})\approx 3$, and Larmor timescale $\rho_{\rm s}/(c\Delta t)\approx 36$. The corresponding synchrotron cooling time is of the order of $t^{\rm LC}_{\rm syn}\sim 3m_{\rm e}c/(2r^2_{\rm e}\gamma_{\rm s}B^2_{\rm LC})\approx 85\Delta t$.

To prevent spurious numerical effects near the star surface due to strong cooling in the integration of particle trajectories, any particle momentum transverse to the field is removed at each time step within a $r_{\star}$-thick shell around the star. The particle pusher is unchanged everywhere else in the simulation. The radiation reaction force must however be artificially and momentarily reduced in the early stages of the simulation. The transient phase when the magnetosphere is filling up with plasma is quite violent and can generate extreme field strengths and gradients, leading to a crash in the simulation. The simulations are first evolved during one period where the strength of the radiation-reaction force is set to $1\%$ of its nominal value, and it is then slowly increasing in time up to about $10\%$. In a last step, it is gradually increased up to the nominal strength at $t\gtrsim 2P$ when a quasi-steady state can be physically exploited.

An important objective of this work is to study the effect of a rescaling of the magnetic field strength. Indeed, previous PIC simulations used unrealistically low field strength but physical quantities were rescaled to make global simulations doable. Here, we scale the magnetic field strength of the reference simulation down to $B_{\star}=10^4$G, $10^5$G, and $10^6$G, but keep the ratio between all energy scales the same as the $B_{\star}=10^7$G simulation, including the radiation-reaction-limited lepton Lorentz factor $\gamma_{\rm rad}$, given in Eq.~(\ref{eq::scale_separation}). To keep the $\gamma_{\rm pc}/\gamma_{\rm rad}$ ratio identical to the $B_{\star}=10^7$G run, the strength of the radiation reaction force --or equivalently the Thomson cross section-- must be boosted by a factor $f_{\rm rad}=10^9$ for $B_{\star}=10^4$G, $f_{\rm rad}=10^6$ for $B_{\star}=10^5$, and $f_{\rm rad}=10^3$ for $B_{\star}=10^6$G. Table~\ref{table::sim_params} lists the numerical and physical parameters used in this study for the reference simulation.

\subsection{Results}\label{sect::results}

Figure~\ref{fig::densities} shows the overall magnetospheric structure achieved by the hybrid model, once the initial transient has left the box and the full scale radiation-reaction force is established at $t\approx 2.5 P$ for the fiducial full-scale $B_{\star}=10^7$G millisecond pulsar. The solution is consistent with previous global PIC studies, it is characterized by prominent separatrices and reconnecting current layers whose base stands near the light cylinder (the Y-point). It is important to note that this solution has not completely reached a steady state. During the initial transient phase, the Y-point forms well inside the light cylinder (around $0.6 R_{\rm LC}$), and slowly migrates towards $R_{\rm LC}$, and still continues to do so at this stage. This migration is still visible in Figure~\ref{fig::densities} in the form of density striations inside the Y-point, a feature that should vanish if a perfect steady state is achieved. The separatrix current layers are also quite thick and have not fully relaxed to the expected thin rim along the last closed field lines, and thus the final state may depart from what is shown here. Nevertheless, the magnetospheric structure in the wind zone where the equatorial current forms, which is the region of prime interest for what follows, has reached the desired state. The reconnecting layer is filled with a chain of plasmoids of various sizes, ranging from layer-thickness scales to stellar-radii scales for the largest ones. The layer is also slightly kinked due to current-driven instabilities as also reported in previous studies (e.g., see \citealt{2015MNRAS.448..606C, 2015MNRAS.449.2759B}). Large vacuum gaps form outside the layer, a feature also already reported in previous studies \citep{2014ApJ...795L..22C, 2015ApJ...815L..19P}. The transition layer in the blending function (Eq.~\ref{eq::transition}) allows to bridge the pure force-free region to these vacuum gaps without noticeable numerical artefacts. We have also verified that this feature is robust against the latitudinal extent of the PIC domain. While plasma depletion is expected in this regime, it is probably exaggerated here because of our pair production scheme, where the pair producing photons have virtually zero mean-free path. However, we believe that this has no dramatic effect on the high-energy output of the magnetosphere described below. 

Perhaps the most spectacular feature is the extreme narrowness of the current layer, the latter being barely visible in the global view (see the zoomed-in view in Figure~\ref{fig::densities}). Near the light cylinder, the layer thickness is about $\delta\sim 10^{-3}R_{\rm LC}$ which corresponds to the scale separation anticipated in Sect.~\ref{sect::scales}, indicating that the simulation has achieved the desired scale separation. The pair multiplicity in the layer is $\kappa\sim 10-100$, while the proton density is at most of the order unity. The pairs being much more abundant than the protons, the layer thickness is controlled by the electronic scales. In contrast, the proton beam is much thicker, with a width $\sim 0.1 R_{\rm LC}$ near the light-cylinder, and it is not significantly perturbed by the presence of plasmoids. The decoupling between the proton and the pair scales is amplified by radiative cooling, for which the protons are nearly immune from.

\begin{figure}
    \centering
    \includegraphics[width=\hsize]{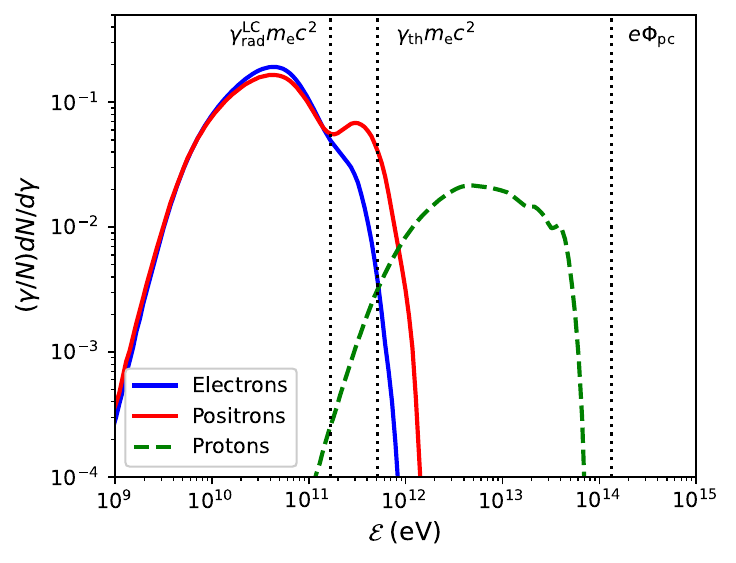}
    \caption{Total electron (solid blue), positron (solid red), and proton (dashed green) energy spectra for a $B_{\star}=10^7$G and $P=1$ms pulsar. The relevant energy scales introduced in Sect.~\ref{sect::scales} are shown by vertical dotted lines, from the lowest to the highest energies: the radiation-reaction-limited energy at the light cylinder, $\gamma^{\rm LC}_{\rm rad}m_{\rm e}c^2$, the threshold energy for pair production, $\gamma_{\rm th}m_{\rm e}c^2$, and the polar-cap potential drop, $e\Phi_{\rm pc}$.}
    \label{fig::spectra}
\end{figure}

\begin{figure}
    \centering
    \includegraphics[width=\hsize]{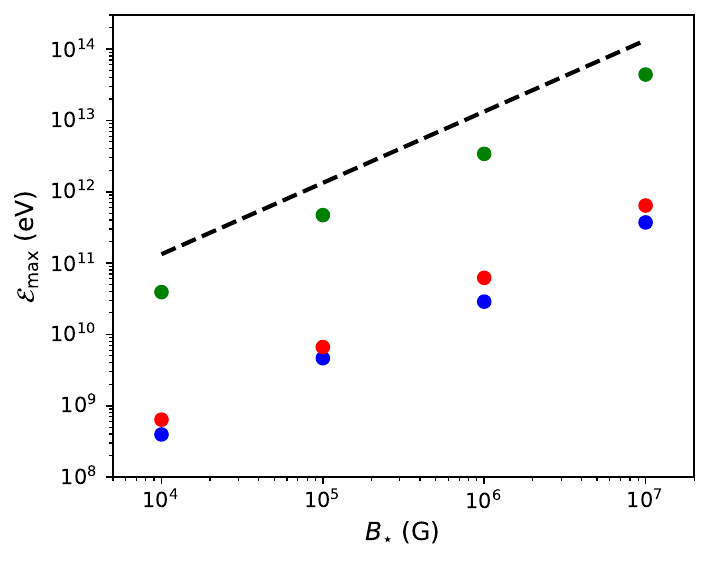}
    \caption{Maximum cut-off energy of the charged particle energy distributions, $\mathcal{E}_{\rm max}$ (electrons: blue dots, positrons: red dots, protons: green dots), as a function of the surface magnetic field $B_{\star}$. The evolution of the polar-cap potential drop is shown by the dashed line for comparison (Eq.~\ref{eq::gamma_pc}).}
    \label{fig::scaling}
\end{figure}

Figure~\ref{fig::densities} also maps out the mean particle energies in the magnetosphere, demonstrating the crucial role of reconnection in particle acceleration. X-points, where the field reconnects in-between two plasmoids, are the main acceleration sites for the pairs. An examination of the bulk velocities reveals that electrons and positrons form two counter-streaming beams at X-points. Positrons (and protons) move radially outwards along with the wind and plasmoids, while electrons tend to precipitate inwards. Counter-streaming is needed to carry the magnetospheric return current \citep{2015MNRAS.448..606C, 2020MNRAS.491.5579C}; it is strongest near the light cylinder and tends to diminish with distance. This effect could also be reduced by a larger pair multiplicity in the current sheet. The zoomed-in view on one of this X-point present in the layer (see the inset in Figure~\ref{fig::densities}) shows that the energy of the pairs peaks inside the layer, $\mathcal{E}_{\rm pairs}=\gamma m_{\rm e}c^2\approx 1$TeV, and sharply decreases inside the nearest plasmoids on either side, down to $\mathcal{E}_{\rm pairs}\lesssim 100$GeV. In this extreme synchrotron cooling regime, particles accelerate deep inside the layer where the effective magnetic field strength is low, allowing them to accelerate beyond the radiation-reaction-limited energy, $\gamma^{\rm LC}_{\rm rad}m_{\rm e}c^2\approx 150$GeV$\ll \mathcal{E}_{\rm pairs}$; but as soon as they leave the layer and enter a plasmoid, they feel a sharp increase of the perpendicular magnetic field, they cool off catastrophically and radiate all their energy away in the form of synchrotron photons \citep{2013ApJ...770..147C}. Plasmoids effectively act as beam dumps, as clearly visible in Figure~\ref{fig::densities}. Thus, other acceleration processes involving, for instance, plasmoid contraction or particle escape from plasmoids are irrelevant for the pairs in this environment, in contrast to non-radiative reconnection \citep{2018MNRAS.481.5687P, 2021ApJ...922..261Z}. However, plasmoids, X-points, and radiative cooling are unimportant for ions. From their perspective, the layer is a sharp featureless magnetic null along which they follow relativistic Speiser orbits \citep{1965JGR....70.4219S, 2007A&A...472..219C, 2013ApJ...770..147C, 2019MNRAS.487..952C, 2021ApJ...922..261Z, 2023ApJ...959..122C}, allowing them to reach extremely-high energies, $\mathcal{E}\approx 10$TeV, close to the full polar-cap potential drop \citep{2018ApJ...855...94P, 2020A&A...635A.138G}. This energy scale is consistent with system-size-limited acceleration achievable with reconnection, given by $\mathcal{E}_{\rm max}\sim \beta_{\rm rec} e \Phi_{\rm pc}\approx \beta_{\rm rec}\times 130$TeV, where $\beta_{\rm rec}\sim 0.1$-0.2 is the reconnection rate.

Figure~\ref{fig::spectra} shows the energy distribution for all the charged particle species. The pair spectrum ranges from about 1GeV to 1TeV and peaks at about a few tens of GeV, below the radiation-reaction limited energy ($\gamma^{\rm LC}_{\rm rad}m_{\rm e}c^2\approx 150$GeV). While there is no major difference between electrons and positrons at low energies, positrons significantly dominate over the electrons at the highest energies. Positrons clearly show an extra spectral component peaking 300-400GeV, which coincides with particles energized at X-points whose energy is determined by the magnetization parameter $\sigma_{\rm LC}m_{\rm e}c^2\approx 500$GeV (Eq.~\ref{eq::sigmalc}). This feature was already reported in previous works \citep{2015MNRAS.448..606C}, and seems to hold strong at more realistic scales. This asymmetry is due to the overall polarization of the magnetosphere where a positive net charge is expected in the equatorial regions, if $\boldsymbol{\Omega}\cdot\boldsymbol{\mu}>0$. Note that this effect is strongest for an aligned rotator and vanishes for an orthogonal rotator \citep{2018ApJ...855...94P}. As anticipated in Sect.~\ref{sect::scales}, the pair spectrum is relatively narrow in energy band for a weak millisecond gamma-ray pulsar because of the proximity between the relevant energy scales. The proton spectrum is composed of a single component ranging from about 1TeV up to 30TeV. If we venture to fit a power law, the spectral index would be $\sim -1.5$, in qualitative agreement with local reconnection studies \citep{2014A&A...570A.112M, 2018MNRAS.473.4840W}. Although the exact spectral shape of all species must be considered with caution, given the oversimplified model used for pair production in this work, we believe that the energy scales, width, and budget are robust overall. Scaling down the magnetic field strength, and keeping all energy scale separations the same with respect to the polar-cap potential drop $\gamma_{\rm pc}$, does not change the above picture. All magnetospheric features are recovered and the particle maximum energies scale linearly with the field strength (Figure~\ref{fig::scaling}). This result suggests that the rescaling performed in previous studies is legitimate as long as the hierarchy of scales of the targeted system is respected.

\begin{figure}
    \centering
    \includegraphics[width=\hsize]{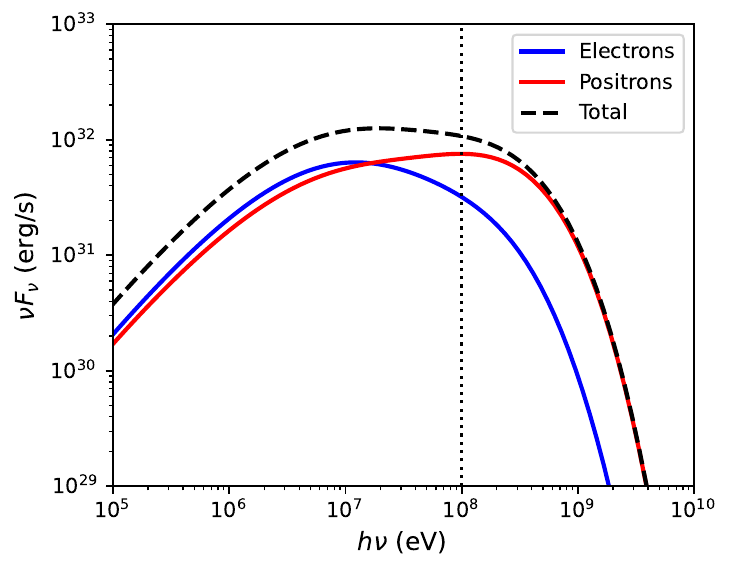}
    \caption{Total synchrotron spectral energy distribution emitted by the equatorial current sheet for $B_{\star}=10^7$G and $P=1$ms (black dashed line). The electronic (positronic) contribution is shown by the blue (red) line. The vertical dotted line indicates the 100MeV energy threshold of the \emph{Fermi}-LAT for comparison.}
    \label{fig::synchrotron}
\end{figure}

\begin{figure}
    \centering
    \includegraphics[width=\hsize]{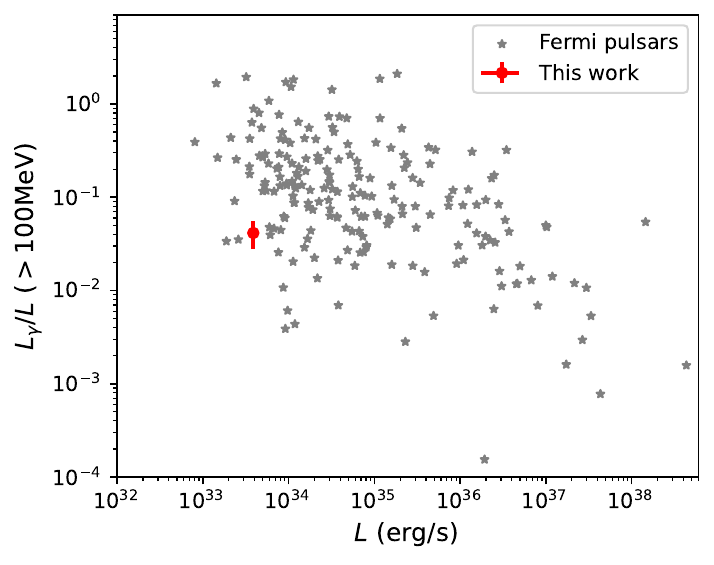}
    \caption{Pulsar gamma-ray efficiency above 100MeV, that is the gamma-ray luminosity over the pulsar spindown power ($L_{\gamma}/L$), as reported in the third \emph{Fermi}-LAT catalogue (gray stars, \citealt{2023ApJ...958..191S}), as a function of the pulsar spindown power ($L$). The properties of the pulsar simulated in this work ($B_{\star}=10^7$G, $P=1$ms) are shown by the red dot. The error bars give a measure of the variations of the spindown and radiative efficiencies measured in a $0.5P$ time frame.}
    \label{fig::efficiencies}
\end{figure}

The narrowness of the pair energy spectrum translates into a narrow-band synchrotron energy distribution. The total synchrotron spectrum reported in Figure~\ref{fig::synchrotron} ranges from about 1MeV to 1GeV, with a peak in the energy band of the \emph{Fermi}-LAT above 100MeV. It includes the emission from the equatorial current layer only, and thus synchrotron radiation largely dominates over curvature radiation because the Larmor radius is much smaller than the particle guiding center curvature radius ($\gtrsim R_{\rm LC}$). Given that the emission is strongly beamed, it corresponds to the spectrum an observer would see if looking in the equatorial plane of the system. The GeV component corresponds to the particles (mostly positrons) accelerated at X-points beyond the synchrotron burnoff limit, given by $\epsilon_{\rm rad}\sim (9/4)m_{\rm e}c^2/\alpha\approx 160$MeV \citep{2011ApJ...737L..40U}, where $\alpha$ is the fine structure constant. Integrating over all frequencies, the synchrotron luminosity accounts for about $18\%$ of the pulsar spindown. This sizeable fraction corresponds to the amount of dissipation of the Poynting flux measured in the magnetosphere within a few light-cylinder radii away from the base of the layer that is controlled by the reconnection rate \citep{2020A&A...642A.204C, 2023ApJ...943..105H}. It is way larger than the numerical dissipation reported in Sect.~\ref{sect::monopole} ($\lesssim 0.1\%$), thus demonstrating that dissipation is by far physical and due to magnetic reconnection. This also means that nearly all of the dissipated power in the box is channelled into non-thermal radiation, which is possible only in the strong cooling regime. Indeed, about $2\%$ of the spindown is channelled into the beam of protons while the pairs leave the box with $\lesssim 1\%$ of the spindown power. The energy fraction carried by the pairs and the protons would increase at larger distances where dissipation proceeds and synchrotron cooling becomes inefficient. Integrated above the energy threshold of the \emph{Fermi}-LAT, our fiducial pulsar shines about $4\%$ of its total spindown power above $\>100$MeV range. Compared to the \emph{Fermi} pulsar population, our solution is compatible with the typical reported gamma-ray efficiencies that range from 1-100\% (\citealt{2023ApJ...958..191S}, see Figure~\ref{fig::efficiencies}).

\section{Conclusion}\label{sect::conclusion}

In this work, we have developed a new hybrid scheme designed to blend the PIC and the force-free approaches in the context of relativistic magnetospheres. This effort is motivated by the need to increase the separation between the kinetic scales where particle acceleration operates and global scales, and by this means add more credibility to global PIC models in spite of their unrealistically low scale separations. Using these new capabilities and high-resolution simulations, we focused our numerical resources on the modeling of a weak millisecond gamma-ray pulsar using realistic (i.e., without artificial rescaling) energy scales that are relevant in the equatorial current sheet, while sacrificing the polar cap physics near the star surface. This hybrid approach is about four times faster than a full PIC simulation of the same problem with identical numerical parameters. Perhaps more importantly, it allows to circumvent the severe resolution constraints imposed by the physical conditions at the polar caps where the plasma is not as energetic as in the current layers, which would prevent a full PIC simulation to model such system with the same numerical resolution as the hybrid run.

Our results recover all features reported by previous studies: a prominent reconnecting current layer efficiently accelerating pairs and ions. For the surface magnetic field $B_{\star}=10^7$G simulated in this work, pairs are accelerated up to 1TeV, while protons reach up to $\gtrsim 10\%$ of the full polar cap potential drop, that is $\gtrsim 10$TeV. With a luminosity of a few percent of the pulsar spindown power, the millisecond pulsar population could thus be a sizeable contributor to the Galactic cosmic-ray population \citep{2018ApJ...855...94P, 2020A&A...635A.138G}. The extreme cooling rate of pairs leads to a nearly complete conversion of the pair energy into synchrotron radiation, including above the burnoff limit in the GeV range. The bolometric radiative efficiency reaches about $20\%$ of the spindown power, including about $4\%$ above 100MeV gamma rays, a number compatible with the \emph{Fermi}-LAT pulsar population \citep{2023ApJ...958..191S}. The fact that pairs can be accelerated at least up to TeV energies, even for a weak pulsar, is encouraging for explaining the recent detections of pulsed TeV emission in the Crab and the Vela pulsars \citep{2016A&A...585A.133A, 2023NatAs...7.1341H}.

We have also shown that the rescaling procedure operated by previous PIC studies, in particular in the surface magnetic field strength, is valid as long as a realistic hierarchy of scales is respected with regard to the targeted system. Thus, the results from this simulation may not be extrapolated to larger field strength, as for instance the young pulsar population, because the scale separation would be different for a larger surface field strength. This work is meant to provide a proof of concept that magnetic reconnection within the equatorial current sheet is a viable scenario to explain the gamma-ray emission, without scaling down the physical parameters, at least in the low-luminosity range of the gamma-ray pulsar population. In this respect, our fiducial simulation has successfully achieved this objective, although much remains to be done in the future. In particular, including a more realistic model for pair production could significantly improve the predictive power of the simulations, in particular regarding the pair multiplicities in the magnetosphere, the particle/photon spectral shapes and cutoff energies, as well as the radiative efficiencies. These last two quantities reported in this work being low compared to the bulk of the \emph{Fermi}-LAT millisecond pulsar population. Longer integration times to establish a more robust steady state, in particular inside the light cylinder (separatrices and Y-point), would be desirable. The hybrid method presented in this paper provides interesting new avenues to model relativistic magnetospheres, including those forming around black holes. More work is needed to apply this method to a more diverse array of astrophysical problems, and to generalize it into a full 3D setup.

\begin{acknowledgements}
We thank Kyle Parfrey and Jens Mahlmann for their valuable comments and encouragements during the implementation of this new numerical scheme. This project has received funding from the European Research Council (ERC) under the European Union’s Horizon 2020 research and innovation programme (grant agreement No 863412). Computing resources were provided by TGCC under the allocation A0130407669 made by GENCI. We thank the International Space Science Institute (ISSI) for providing financial support for the organization of the meeting of ISSI Team No 459 led by I. Contopoulos and D. Kazanas where this project has been discussed. Software: Matplotlib (\citealt{Hunter:2007}), Numpy (\citealt{harris2020array}).
\end{acknowledgements}

\bibliographystyle{aa}
\bibliography{references.bib}

\end{document}